\documentclass[preprint,12pt]{elsarticle}

\usepackage{amssymb}
\usepackage{amsmath}
\usepackage[version=4]{mhchem}

\usepackage{tikz}
\DeclareRobustCommand\full  {\tikz[baseline=-0.6ex]\draw[thick] (0,0)--(0.5,0);}
\DeclareRobustCommand\dotted{\tikz[baseline=-0.6ex]\draw[thick,dotted] (0,0)--(0.54,0);}

\DeclareRobustCommand\dashdotted {\tikz[baseline=-0.6ex]\draw[thick,dash dot] (0,0)--(0.5,0);}

\usepackage{flafter}
\usepackage{placeins}

\usepackage[nolist,nohyperlinks]{acronym}

\usepackage{xcolor}
\usepackage[hidelinks]{hyperref}

\AddToHook{shipout/firstpage}{%
  \begin{tikzpicture}[remember picture,overlay]
    \node[
      anchor=north west,
      xshift=8mm,
      yshift=-8mm,
      draw,
      fill=white,
      inner sep=4pt,
      text width=0.85\textwidth,
      align=left,
      font=\normalfont\small
    ] at (current page.north west) {%
      \urlstyle{same}%
      This is a pre-print of an article published in:\\
      \textbf{Combustion and Flame}.\\
      The journal version is available online at:\\
      \href{https://doi.org/10.1016/j.combustflame.2023.113206}
           {\textcolor{blue}{\textbf{\nolinkurl{https://doi.org/10.1016/j.combustflame.2023.113206}}}}\\
      Please cite the journal version rather than the arXiv entry.
    };
  \end{tikzpicture}%
}

\journal{Combustion and Flame}

\newcommand{\methane}{CH\textsubscript{4}}
\newcommand{\hydrogen}{H\textsubscript{2}}
\newcommand{\dinitrogen}{N\textsubscript{2}}
\newcommand{\dioxygen}{O\textsubscript{2}}
\newcommand{\water}{H\textsubscript{2}O}
\newcommand{\nox}{NO\textsubscript{x}}
\newcommand{\noo}{NO\textsubscript{2}}
\newcommand{\OHstar}{OH\textsuperscript{*}}

\newcounter{reactionCount} 
\newcommand{\reaction}[1]{
    \refstepcounter{reactionCount} 
    \[ \ce{#1} \tag{R\thereactionCount} \] 
}

\begin{document}

\begin{acronym}
\acro{PAC}{Plasma Assisted Combustion}
\acro{ARC}{Analytically Reduced Chemistry}
\acro{CEMA}{Chemical Explosive Mode Analysis}
\acro{CEM}{Chemical Explosive Mode}
\acro{COG}{Center Of Gravity}
\acro{CSP}{Computational Singular Perturbation}
\acro{NRPD}{Nanosecond Repetitively Pulsed Discharge}
\acro{PRF}{Pulse Repetition Frequency}
\acro{HRR}{Heat Release Rate}
\acro{LES}{Large Eddy Simulation}
\acro{DNS}{Direct Numerical Simulation}
\acro{TF}{Thickened Flame}
\acro{OP}{Operating Point}
\acro{NEP}{Non Equilibrium Plasma}
\acro{QSS}{Quasi-Steady State}
\acro{DRGEP}{Directed Relation Graph with Error Propagation}
\acro{3D}{3-Dimensional}
\acro{2D}{2-Dimensional}
\acro{1D}{1-Dimensional}
\acro{0D}{0-Dimensional}
\acro{PSR}{Perfectly Stirred Reactors}
\acro{LOS}{Line-of-Sight}
\acro{COG}{Center of Gravity}
\acro{RHS}{Right-Hand Side}
\acro{SB}{Sequential Burner}
\acro{EEDF}{Electron Energy Distribution Function}
\end{acronym}

\begin{frontmatter}



\title{Numerical study of nitrogen oxides chemistry during plasma assisted combustion in a sequential combustor}


\author[inst1]{Quentin Malé\corref{cor1}}\ead{qumale@ethz.ch}

\affiliation[inst1]{organization={CAPS Laboratory, Department of Mechanical and Process Engineering, ETH Zürich, 8092 Zürich, Switzerland}} 

\author[inst2]{Nicolas Barléon}

\affiliation[inst2]{organization={CERFACS, 42 avenue Gaspard Coriolis, 31057 Toulouse, France}} 

\author[inst1]{Sergey Shcherbanev}
\author[inst1]{Bayu Dharmaputra}
\author[inst1]{Nicolas Noiray\corref{cor1}}\ead{noirayn@ethz.ch}

\cortext[cor1]{Corresponding authors}

\begin{abstract}
Plasma Assisted Combustion (PAC) is a promising technology to enhance the combustion of lean mixtures prone to instabilities and flame blow-off. Although many PAC experiments demonstrated combustion enhancement, several studies report an increase in \nox{} emissions. 
The aim of this study is to determine the kinetic pathways leading to \nox{} formation in the second stage of a sequential combustor assisted by Nanosecond Repetitively Pulsed Discharges (NRPDs). 
For this purpose, Large Eddy Simulation (LES) associated with an accurate description of the combustion/\nox{} chemistry and a phenomenological model of the plasma kinetics is used. Detailed kinetics 0-Dimensional reactors complement the study. First, the LES setup is validated by comparison with experiments. Then, the \nox{} chemistry is analyzed. 
For the conditions of operation studied, it is shown that the production of atomic nitrogen in the plasma by direct electron impact on nitrogen molecules increases the formation of NO. Then, the NO molecules are transported through the turbulent flame without being strongly affected. 
This study illustrates the need to limit the diatomic nitrogen dissociation process in order to mitigate harmful emissions. More generally, the very good agreement with experimental measurements demonstrates the capability of LES combined with accurate models to predict the NRPD effects on both turbulent combustion and \nox{} emissions.

\end{abstract}




\end{frontmatter}



\section*{Novelty and Significance Statement}
In this work, a novel high-fidelity Large Eddy Simulation (LES) setup is used for the first time to study the effects of Nanosecond Repetitively Pulsed Discharges (NRPDs) on the formation of \nox{}. The LES setup includes an accurate description of the combustion and \nox{} chemistry with an innovative sophisticated modeling of the plasma discharges to account for the impact on the \nox{} formation. Massively parallel computations of a lab-scale sequential combustor are performed and results are compared with experimental data. Chemical pathways for the formation of \nox{} due to NRPDs are studied for the first time in a real burner. This is of major importance to limit the \nox{} emissions that can occur under certain conditions during plasma assisted combustion. The excellent agreement with the experimental data demonstrates the capability of the original LES setup to retrieve the NRPD effects on the turbulent flame and on the formation of NO.

\section*{Author Contributions}
Q.M. and N.N. conceived the research idea. Q.M. led the numerical simulation investigations. S.S. and B.D. performed the experiments. N.B. and Q.M. set up the plasma modeling framework. All authors discussed the results. The manuscript was written by Q.M. All authors reviewed and approved the final version of the manuscript. 

%

\section{Introduction}
\label{sec:introduction}


\acf{PAC} is being actively developed to improve combustion systems efficiency and to reduce harmful emissions. In particular, \ac{PAC} can help to control ultra lean combustion, to develop low \nox{} emissions hydrogen combustion technologies, or to control the combustion of alternative fuels \cite{Ju2015a}. \acfp{NRPD} have been successfully applied to enhance the combustion in various combustion systems. A recent state of the art has been presented in Ref.~\cite{Lacoste2022}. For example, \acp{NRPD} have been shown to significantly extend lean combustion limits (e.g., \cite{Pilla2008,Barbosa2015,Xiong2019,DiSabatino2020,Vignat2021,Shcherbanev2022,Male2022}), suppress unstable combustion regimes (e.g., \cite{Lacoste2013,Moeck2013}) and improve flame anchoring (e.g., \cite{Shcherbanev2020,DISabatino2021}), in combustors relevant for gas turbine applications.

Although the results to date are very promising, some studies report an increase in \nox{} emissions when plasma discharges are used \cite{Kim2007,Lacoste2013a,Kim2017a,Xiong2019}, without investigating in detail the origin of these emissions. This could be due to the interaction of the plasma with the N\textsubscript{2} molecules. The plasma can, for example, dissociate \dinitrogen{} by direct electron impact 
\reaction{e^- + N2 -> e^- + N + N \text{.} \label{ceq:e-N2dissoc}}
The N atom can then be responsible for NO production through the reactions 
\reaction{N + O2 <=> NO + O \text{,} \label{ceq:N+O2}} 
\reaction{N + OH <=> NO + H \, \text{and} \label{ceq:N+OH}} 
\reaction{N + O + M <=> NO + M \text{.} \label{ceq:N+O}} 
Other species can also act on the \nox{} formation. For example, atomic oxygen is known to be generated by the dissociative quenching of electronically excited diatomic nitrogen N\textsubscript{2}\textsuperscript{*} with \dioxygen{} \cite{Popov2001}
\reaction{N2^* + O_2 <=> 2 O + N2 \text{,} \label{ceq:N2*+O2}} 
and it can then react with N\textsubscript{2} to form NO through
\reaction{N2 + O <=> NO + N \text{.} \label{ceq:N2+O}} 
The atomic N from Eq.~(\ref{ceq:N2+O}) can then trigger Eq.~(\ref{ceq:N+O}). Furthermore, plasma discharges in \water{}-containing mixtures can generate OH molecules that will set off Eq.~(\ref{ceq:N+OH}). Plasma discharges are therefore a potential source of \nox{}. However, it is also known that the flame following the plasma can consume nitrogen oxides, similar to a reburn process \cite{Kim2007}. 
The study of the \nox{} chemical kinetics during \ac{PAC} is therefore extremely complex because of the coupling between the plasma, combustion, and \nox{} chemistry. 

In this work, \ac{LES}\textemdash supported by experiments\textemdash is used as a tool to investigate the formation of nitrogen oxides in the second stage of a sequential combustor assisted by \acp{NRPD} where:
\begin{itemize}
    \item[-] The sequential burner is fed with a mixture containing combustion products from the first stage flame, including \nox{}.
    \item[-] The burnt gas produced by the first stage undergoes intense cooling and turbulent mixing with cold dilution air and secondary fuel injection.
    \item[-] The mixture of vitiated air and fuel is stratified in the sequential burner, both in terms of composition and temperature.
    \item[-] The plasma induces an excitation of the species flowing between the electrode system, resulting in rapid heating, slow heating, and the generation of active species in the reactive flowing mixture.
    \item[-] The turbulent combustion process that occurs in the second stage is affected by the plasma generated upstream.
\end{itemize}
The objective of this study is to clarify the mechanisms responsible for the formation of \nox{} in this highly complex environment. Only a massively parallel \ac{LES} of the full sequential stage can address this properly. From a numerical point of view, this involves multiple challenges:
\begin{itemize}
    \item[-] The multitude of physical processes considered (i.e., turbulent mixing, turbulent combustion, \nox{} chemistry, plasma chemistry, combustion chemistry) requires a powerful and scalable multi-physics computation code.
    \item[-] The interaction of \nox{}, plasma and combustion chemistry demands a sophisticated description of the chemical kinetics.
    \item[-] The computational cost associated with the spatial and temporal resolution of the plasma discharges including the description of the electric field in a \ac{3D} high Reynolds flow is prohibitive. A model capturing all important plasma effects must be derived to enable \ac{3D} turbulent reactive flow simulation in realistic geometries.
    \item[-] Even simplified, plasma effects occur on a very small time scale both in time and space compared to combustion processes. The integration of the plasma requires very small grid cells ($\Delta_x \approx 100$~$\mathrm{\mu m}$) and very small time steps ($\Delta_t \approx 0.1$~ns) which increases the computational cost considerably.
    \item[-] In addition to a plasma model and a reasonably complete description of chemistry, a proper method to model subgrid scale turbulence-chemistry interaction is also needed for the turbulent flame of the second stage combustion chamber.
\end{itemize}

This paper shows how sophisticated/high performance simulation tools combined with experiments can be used to advance recent combustion technology by gaining knowledge about the physical phenomena involved. 
First, the configuration of the sequential combustor and the operating points chosen for this work are described. 
Then, the simulation and modeling methods set up are detailed. 
Finally, the results of the simulations are analyzed to study the \nox{} formation mechanisms. The \ac{LES} data are supported by experimental measurements.

\section{Sequential combustor configuration}
\label{sec:seqcombconfig}

\begin{figure}[hbt!]
    \centering
    \includegraphics[width=1.0\textwidth]{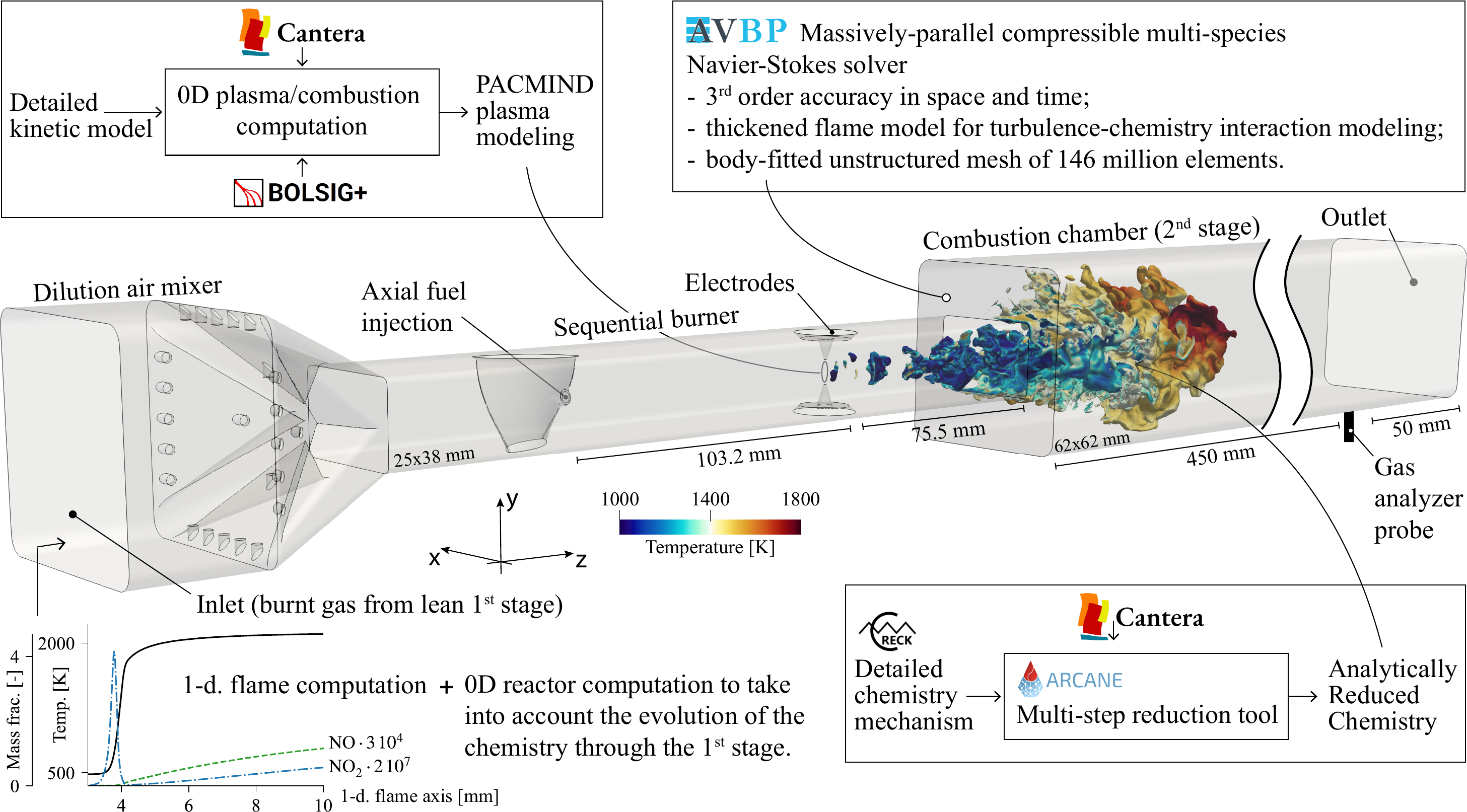}
    \caption{Diagram of the sequential stage with the detail of its geometry and its components, as well as the illustration of the numerical tools used to perform the \acp{LES} including the plasma effects and the \nox{} chemistry. The temperature colored isosurface of heat release rate depicts the flame assisted by plasma in the sequential stage.}
    \label{fig:sketch_LES_seqcomb}
\end{figure}

This numerical study is based on experimental tests performed with a laboratory scale sequential combustor operated at atmospheric pressure. The first stage consists of an array of 4$\times$4 technically premixed jet flames, which generate hot combustion products. The focus of this work is on the sequential stage where \acp{NRPD} are applied. Therefore, only this stage is simulated. Figure~\ref{fig:sketch_LES_seqcomb} shows a diagram of the sequential stage together with the simulation and modeling tools implemented for the study. The sequential stage features a dilution air mixer with large lateral vortex generators and multiple air injection holes for rapid mixing, a \acf{SB} in which fuel is injected axially and mixes with the co-flow of hot vitiated air, a pin-to-pin electrodes arrangement in the \ac{SB} which enables \ac{NRPD} generation upstream of the burner outlet, and the sequential combustion chamber. The inlet composition and temperature are defined by the combustion products of the first stage: a \ac{1D} laminar flame is used to estimate the combustion products of the first stage flame, then, the mixture evolves in a \ac{0D} reactor for a time equal to the residence time until the inlet of the sequential stage dilution air mixer. The validity of this approach is assessed by comparing the NO concentration level without second stage fuel injection between the experiment and the simulation (``2\textsuperscript{nd} stage off'' in Fig.~\ref{fig:NOX_values}).

Throughout this study, the thermochemical state of the vitiated air flow that feeds the sequential stage is fixed. The first stage flue gas results from the combustion of $15$~g/s of air with $0.7$~g/s of methane at $503.15$~K, the equivalence ratio is $0.8$. It is then diluted with $18$~g/s of air at ambient temperature. A \methane{}/\hydrogen{} fuel blend is used in the sequential stage. The mass flows of methane and hydrogen are $0.6$~g/s and $0.07$~g/s, respectively. The global equivalence ratio in the sequential stage is $0.6$, the flame thermal power is $38.4$~kW.

High voltage pulses are applied at a certain \ac{PRF} to locally generate multiple \acp{NRPD} between the electrodes. All operating points are summarized in Table~\ref{tab:parameters}, including details on plasma discharges where applicable. $E^p$ is the pulse deposited energy measured experimentally. $P^p$ is the mean plasma power defined as
\begin{equation}
    P^p = E^p \, \mathrm{PRF} \, \mathrm{.}
\end{equation}

\begin{table}[hbt!] \small
\centerline{\begin{tabular}{lccccc}
\hline
Label    & Fuel inj. & Voltage & $E^p$ & PRF & $P^p$ \\
\hline
2\textsuperscript{nd} stage off & no & - & - & - & - \\
NRPD off & yes & - & - & - & - \\
$14$~kV, $20$~kHz & yes & 14~kV& $1.21$ mJ& 20~kHz& $24.2$~W\\
$14$~kV, $40$~kHz & yes & 14~kV& $2.07$ mJ& 40~kHz& $82.8$~W\\
\hline
\end{tabular}}
\caption{Varied parameters for the different \acfp{OP} studied in the sequential stage. The labels will be used later to refer to the different \acp{OP}.}
\label{tab:parameters}
\end{table}

In the experiments, the reacting flow in the 2\textsuperscript{nd} stage of the sequential combustor is characterized by recording $\mathrm{OH}^{*}$ chemiluminescence. The detection system consists of a high-speed CMOS camera (LaVison Star X) coupled to a high-speed intensifier (LaVision HS-IRO), equipped with a CERCO UV lens ($45$~mm, $\mathrm{F/1.8}$) and a bandpass filter (Edmund Optics, centered at $310$~nm, FWHM $10$~nm). The camera frame rate is set to 5~kHz with an intensifier gate width of 40~$\mathrm{\mu}$s. The emission from the discharges is much more intense than the $\mathrm{OH}^{*}$ emission of the flame. An optical obstacle is therefore used to mask the discharge location and to prevent damage to the IRO-intensifier.

Characterization of the discharge geometry was performed during a previous experimental campaign 
\cite{Male2022}. 
The volume of the discharge is estimated as a cylinder with a radius of $500$~$\mathrm{\mu m}$ and a length of $5$~mm.
These measurements were obtained from an average of 100 image samples of the plasma channel. The images were acquired with a camera equipped with a narrow band $337\pm$5~nm filter, capturing the most intense band of N\textsubscript{2}(C-B)(0-0) transition.

The exhaust gas analysis is performed using an ABB EL3040 gas analyzer (sampling rate 1 Hz) with an Uras26 infrared photometer. The device has a $7.9$\% relative extended uncertainty on NO measurements in the range $0\-200$~$\mathrm{mg/Nm^3}$. The sampling is performed for $30$~s of operation near the outlet of the sequential combustor, $450$~mm after the inlet of the 2\textsuperscript{nd} stage combustion chamber (Fig.~\ref{fig:sketch_LES_seqcomb}), through a $2$~mm pinhole of a gas probe. The dry NO emissions are reported in terms of $\mathrm{mg/Nm^3}$, conditioned at $276.15$~K.

\section{Numerical model and methods}
\label{sec:numericalmodelmethods}

\subsection{Plasma modeling}
\label{sec:plasma_modeling}

During each high-voltage nanosecond pulse, the kinetic energy of the electrons increases. Subsequently, this energy is transferred to the surrounding heavy particles through collisional processes, giving rise to excited particles, ions and radicals. Despite the recent efforts to reduce \ac{NEP} chemical kinetics \cite{Bellemans2020,Bellemans2020c,Cheng2022}, the computational cost of solving the electric field, the electron energy, the continuity equations for charged and excited species, in addition to the Navier-Stokes equations for reactive flows is very high. The integration of \ac{NEP} effects in \ac{3D} simulations of turbulent reactive flows inside complex geometries can only be achieved with simplified models that incorporate all relevant plasma effects for the considered study. 
Such a model has been developed by Castela et al. \cite{Castela2016}, incorporating the most important effects of \ac{NEP} in a N\textsubscript{2}/O\textsubscript{2} mixture. In this model, the relaxation of electronically excited N\textsubscript{2} molecules translates into a fast increase of the gas temperature and into the fast dissociation of O\textsubscript{2} molecules, whereas the relaxation of vibrationally excited N\textsubscript{2} molecules translates into a slower heat release. 
The model has been successfully applied to analyze the flow induced by a plasma discharge in a \ac{DNS} framework \cite{Castela2017}. It has also been successfully applied in a \ac{LES} framework to study turbulent flame ignition \cite{Bechane2020}, turbulent combustion enhancement \cite{Bechane2022} and stabilization \cite{Male2022}.
Although this simple model has proven to be very effective, it does not account for secondary effects that may be responsible for the production of \nox{}. A more complex model must therefore be used for the present study. Recently, Barléon et al. \cite{Barleon2023} have extended the phenomenological approach of Castela et al. \cite{Castela2016} to take into account more plasma processes involved in methane-air mixtures, including the production of N and NO, which is crucial for the study of \nox{} formation. Therefore, this approach will be used in this work. As in Ref.~\cite{Castela2016}, the discharge power $\dot{E}^p$ is divided into three contributions:
\begin{equation}
    \dot{E}^p = \dot{E}_{\mathrm{chem}}^p + \dot{E}_{\mathrm{heat}}^p + \dot{E}_{\mathrm{vib}}^p \, \text{,}
    \label{eq:discharge_total_power}
\end{equation}
where $\dot{E}_{\mathrm{chem}}^p$, $\dot{E}_{\mathrm{heat}}^p$ and $\dot{E}_{\mathrm{vib}}^p$ are the amount of power that goes into chemical effects, fast gas heating and vibrational excitation, respectively. However, $\dot{E}_{\mathrm{chem}}^p$ is here divided into several global processes instead of a single one in the original work of Castela et al. \cite{Castela2016}. This allows to consider a wide variety of chemical processes occurring during the discharge. The amount of power that goes into chemical effects reads
\begin{equation}
    \dot{E}_{\mathrm{chem}}^p = \sum_{j=1}^P \alpha_j \dot{E}^p = \alpha_{\mathrm{chem}} \dot{E}^p \, \text{,}
    \label{eq:discharge_chemical_power}
\end{equation}
where $\alpha_j$ is the fraction of the discharge power which contributes to the process $j$ and $\alpha_{\mathrm{chem}} = \sum_{j=1}^P \alpha_j$ is the global chemical energy fraction resulting from $P$ processes. The amount of power that goes into fast gas heating and vibrational excitation read
\begin{equation}
    \dot{E}_{\mathrm{heat}}^p = \alpha_{\mathrm{heat}} \dot{E}^p \; \text{and} \; \dot{E}_{\mathrm{vib}}^p = \alpha_{\mathrm{vib}} \dot{E}^p \, \text{,}
\end{equation}
respectively. The discharge energy fraction coefficients $\alpha$ characterize the discharge effects. These coefficients satisfy:
\begin{equation}
   \sum_{j=1}^P \alpha_j + \alpha_{\mathrm{heat}} + \alpha_{\mathrm{vib}} = 1 \, \text{.}
\end{equation}

The difficulty now lies in determining the $\alpha$ coefficients and the important $P$ processes, that depend on the gas thermochemical state $\mathcal{S} = \left( Y_{k=1,...,N_s}, T, P \right)$ and the electric field $E$, where $N_s$ is the number of species in the gas mixture. A precise methodology has been established by Barléon et al. \cite{Barleon2023} for this purpose. First, detailed kinetic simulations are conducted using isochoric plasma reactors following the approach described in Ref.~\cite{Cheng2022}. These simulations include detailed plasma-combustion kinetics, which consider electron impact reactions, ion chemistry, vibrational and electronic excitation and relaxation. Second, an analysis of the species production rate from plasma processes issuing from the detailed plasma kinetics $\dot{\omega}_k^{p,*}$ determines how the discharge energy is split into chemical, fast heating, and vibrational effects. The $\alpha$ coefficients are ultimately computed to integrate the \ac{NEP} effects into the \ac{LES} equations (Section~\ref{sec:large_eddy_simulation}) where only ground state neutral species are considered. 
A precise description of the method for determining the $\alpha$ coefficients is given in Ref.~\cite{Barleon2023}. It is not reported here for brevity reasons. Table~\ref{tab:PACMIND} gives the global processes used to model the \ac{NEP} discharge effects for the present study, for which \nox{} production must be taken into account and \acp{NRPD} are applied in a mixture of vitiated air and fuel. 
For now, the plasma model neglects the presence of hydrogen molecules in the gas mixture. 
It is verified in Section~\ref{sec:discharge_characterization} that this simplification does not significantly affect the outcome by comparing the modeling results with a detailed plasma-combustion chemistry.



\begin{table}[hbt!]
\centering
\small
\begin{tabular}{|c|c|c|}
\hline   & Dominant physical processes & Global process \\
\hline 1 & $\mathrm{e}^{-}+\mathrm{O}_2 \longrightarrow \mathrm{e}^{-}+2 \mathrm{O}$ & $\mathrm{O}_2 \longrightarrow 2 \mathrm{O}$ \\
& $\mathrm{~N}_2\left(\mathrm{~A}_3, \mathrm{~B}_3, \mathrm{C}_3, \mathrm{a}_1\right)+\mathrm{O}_2 \longrightarrow \mathrm{N}_2+2 \mathrm{O}$ &  \\
\hline 2 & $\mathrm{e}^{-}+\mathrm{N}_2 \longrightarrow \mathrm{e}^{-}+\mathrm{N}+\mathrm{N}\left({ }^2 \mathrm{D}\right)$ & $\mathrm{N}_2 \longrightarrow 2 \mathrm{~N}$  \\
\hline 3 & $\mathrm{~N}\left({ }^2 \mathrm{D}\right)+\mathrm{O}_2 \longrightarrow \mathrm{NO}+\mathrm{O}$ & $\mathrm{O}_2+\mathrm{N}_2 \longrightarrow 2 \mathrm{NO}$  \\
\hline 4 & $\mathrm{e}^{-}+\mathrm{CH}_4 \longrightarrow \mathrm{e}^{-}+\mathrm{CH}_3+\mathrm{H}$ & $\mathrm{CH}_4 \longrightarrow \mathrm{CH}_3+\mathrm{H}$  \\
& $\mathrm{~N}_2\left(\mathrm{~A}_3, \mathrm{~B}_3, \mathrm{C}_3, \mathrm{a}_1\right)+\mathrm{CH}_4 \longrightarrow \mathrm{N}_2+\mathrm{CH}_3+\mathrm{H}$ &  \\
\hline 5 & $\mathrm{e}^{-}+\mathrm{CO}_2 \longrightarrow \mathrm{e}^{-}+\mathrm{CO}+\mathrm{O}$ & $\mathrm{CO}_2 \longrightarrow \mathrm{CO}+\mathrm{O}$  \\
& $\mathrm{~N}_2\left(\mathrm{~A}_3, \mathrm{~B}_3, \mathrm{C}_3, \mathrm{a}_1\right)+\mathrm{CO}_2 \longrightarrow \mathrm{N}_2+\mathrm{CO}+\mathrm{O}$ &  \\
\hline 6 & $\mathrm{e}^{-}+\mathrm{H}_2 \mathrm{O} \longrightarrow \mathrm{e}^{-}+\mathrm{OH}+\mathrm{H}$ & $\mathrm{H}_2 \mathrm{O} \longrightarrow \mathrm{OH}+\mathrm{H}$  \\
& $\mathrm{~N}_2\left(\mathrm{~A}_3, \mathrm{~B}_3, \mathrm{C}_3, \mathrm{a}_1\right)+\mathrm{H}_2 \mathrm{O} \longrightarrow \mathrm{N}_2+\mathrm{OH}+\mathrm{H}$ &  \\
\hline 7 & $\mathrm{CH}_4+\mathrm{O}\left({ }^1 \mathrm{D}\right) \longrightarrow \mathrm{CH}_3+\mathrm{OH}$ & $\mathrm{CH}_4+\frac{1}{2} \mathrm{O}_2 \longrightarrow \mathrm{CH}_3+\mathrm{OH}$  \\
\hline
\end{tabular}
\caption{Description of the global processes considered in this work to model the \ac{NEP} discharge effects.}
\label{tab:PACMIND}
\end{table}

\subsection{Analytically reduced chemistry}
\label{sec:chemistry}

The study of \nox{} chemistry in a sequential combustor with plasma discharges requires a precise description of the chemical kinetics able to account for the evolution of pollutants as well as ignition induced by the \acp{NRPD}, auto-ignition, flame consumption rate and stretch response. \ac{ARC} mechanisms\textemdash whose application to reacting flows is widely described by Felden \cite{Felden:phd2017}\textemdash allow to accurately describe these phenomena, while keeping the computational cost at an affordable level \cite{Jaravel2017,Rochette2018,Male2021,Male2022,Capurso2023}.

The \ac{ARC} mechanism is specially constructed for this work using the ARCANE library \cite{Cazeres2021}, co-developed by CERFACS and Cornell University. ARCANE includes a multi-step reduction tool relying on \ac{DRGEP} \cite{Pepiot-Desjardins2008}, chemical lumping \cite{Pepiot-Desjardins2008a} and \ac{QSS} approximation \cite{Lovas2002}. Reduction starts from the CRECK mechanism \cite{Ranzi2014,Bagheri2020} with an appropriate \nox{} sub-mechanism from Ref.~\cite{Song2019}. The detailed mechanism comprises 159 species and 2459 reactions. 
A set of canonical \ac{1D} or \ac{0D} configurations is used to steer the reduction process towards an accurate ARC mechanism. The configurations include \ac{1D} premixed laminar flames and \ac{0D} homogeneous reactors. 
The fresh gas conditions of the set of 1D flames are representative of the conditions encountered in the sequential stage, with variations of the mixture fraction $Z$ around the perfectly mixed mixture fraction $Z^{PM} = 0.0195$. The mixture fraction is defined here between the fuel ($Z=1$) and the vitiated air after the dilution air mixer ($Z=0$). 
The initial conditions of the set of \ac{0D} reactors are similar.
Nevertheless, additional \ac{0D} reactor configurations are needed to keep the core of the physics of the plasma-induced ignition and \nox{} chemistry. Cases where the initial mixture is doped by an addition of atomic oxygen are created to mimic plasma ignition \cite{Sun2012,Male2022}. Cases where the initial mixture is doped by an addition of NO and \noo{} are created to mimic the effects of \nox{} on the acceleration of \methane{} oxidation \cite{Song2019}. The final \ac{ARC} mechanism contains 39 species (18 of them in quasi-steady state) and 198 reactions (see Supplemental Material).

\begin{figure}[hbt!]
    \centering
    \includegraphics[width=0.45\textwidth]{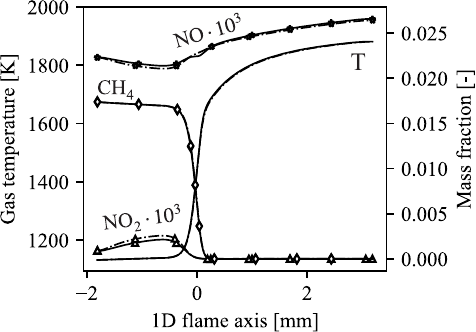}
    \caption{Profiles of temperature and species mass fraction predicted by the reference chemistry (\full{}) and the \ac{ARC} chemistry (\dashdotted{}) for a 1D laminar flame at perfectly mixed conditions, equivalence ratio $\phi = 0.6$.}
    \label{fig:chemistry_1dflame}
\end{figure}

\begin{figure}[hbt!]
    \centering
    \includegraphics[width=0.45\textwidth]{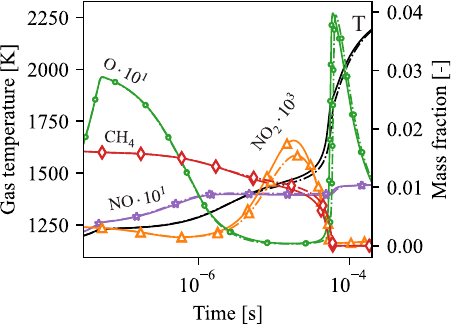}
    \caption{Gas temperature and species mass fraction after a \ac{NEP} discharge modeled using the approach described in Section~\ref{sec:plasma_modeling} for the reference chemistry (\full{}) and the \ac{ARC} chemistry (\dashdotted{}). $E/N = 250$~Td, $e^p = 0.064$~$\mathrm{J/cm^3}$, 
    the initial thermochemical state corresponds to perfectly mixed conditions.}
    \label{fig:chemistry_pacmind_discharge}
\end{figure}

\begin{figure}[hbt!]
    \centering
    \includegraphics[width=0.45\textwidth]{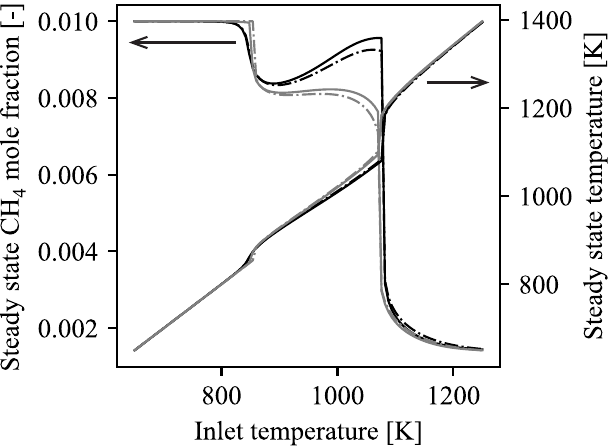}
    \caption{Steady-state methane mole fraction and gas temperature in a \ac{PSR} as defined in Ref.~\cite{Song2019}, function of the inlet temperature for the reference chemistry (\full{}) and the \ac{ARC} chemistry (\dashdotted{}). Black lines: $X_{\mathrm{NO}} = 500$~ppm. Gray lines: $X_{\mathrm{NO_2}} = 400$~ppm. 107~kPa, equivalence ratio $\phi = 2$, diluted with Argon.}
    \label{fig:chemistry_psr_Ysong}
\end{figure}

Figure~\ref{fig:chemistry_flamespeed_ignitiontime} shows the laminar flame speed and the auto-ignition time predicted by the reference chemistry and the \ac{ARC} chemistry for a wide range of mixture fractions, computed with Cantera \cite{cantera}. A good agreement is observed between the detailed and the new \ac{ARC} mechanism. The evolution of the \nox{} through the flame is also consistent (Fig.~\ref{fig:chemistry_1dflame}). 
In order to assess the capability of the \ac{ARC} scheme to capture the plasma effects, a low energy \ac{NEP} discharge $E^p = 0.25$~mJ is applied in a \ac{0D} reactor with a reduced electric field $E/N = 250$~Td representative of the current work (Fig.~\ref{fig:chemistry_pacmind_discharge}). 
The volume of the discharge is experimentally estimated as a cylinder with a radius of $500$~$\mathrm{\mu m}$ and a length of $5$~mm (Section~\ref{sec:seqcombconfig}). This results in an energy density $e^p = 0.064$~$\mathrm{J/cm^3}$. The ignition time and the species evolution are correctly retrieved. 
The enhancement of \methane{} oxidation by NO and \noo{} molecules is assessed using \ac{PSR} at specific conditions for which the mechanism has been validated \cite{Song2019}. The \ac{PSR} computations are performed at $107$~kPa, filled with a mole fraction ratio $\mathrm{CH_4}/\mathrm{O_2}/\mathrm{NO} = 0.01/0.01/0.0005$ and $\mathrm{CH_4}/\mathrm{O_2}/\mathrm{NO_2} = 0.01/0.01/0.0004$, both balanced with Argon (Fig.~\ref{fig:chemistry_psr_Ysong}). This corresponds to the experimental conditions of Ref.~\cite{Song2019}, whose pressure is close to the atmospheric pressure of our burner. The acceleration of the \methane{} oxidation is accurately reproduced. 
These validation cases demonstrate that the chemical pathways of interest for the present work were correctly kept during the reduction process. The behavior of the reduced kinetic system is suitable for the accurate description of plasma assisted combustion and \nox{} chemistry under the conditions of the sequential combustor studied.


\begin{figure}[hbt!]
    \centering
    \includegraphics[width=0.425\textwidth]{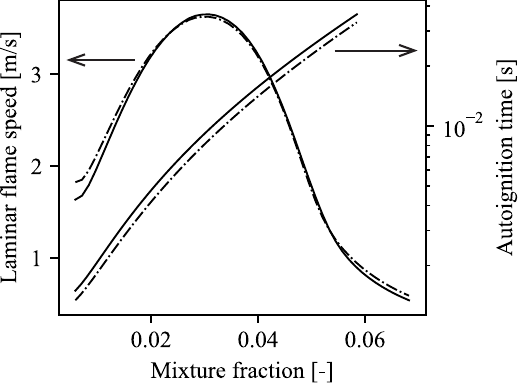}
    \caption{Laminar flame speed and auto-ignition time predicted by the reference chemistry (\full{}) and the \ac{ARC} chemistry (\dashdotted{}). The mixture fraction is defined from the composition of the vitiated air and the fuel.}
    \label{fig:chemistry_flamespeed_ignitiontime}
\end{figure}

\subsection{Large eddy simulation}
\label{sec:large_eddy_simulation}

The \acp{LES} are computed using the AVBP code \cite{Schonfeld1999,Gicquel2011}. AVBP is an unstructured cell-vertex massively-parallel code solving the compressible multi-species Navier-Stokes equations. The discretization of the convective terms is done using the fully explicit Two-step Taylor–Galerkin C (TTGC) finite element numerical scheme \cite{Colin2000a}. TTGC offers third order accuracy in space and time on irregular grids. 
The effects of the unresolved small-scale fluid motions are modeled using the SIGMA \cite{Nicoud2011} subgrid-scale model. 
The so-called \ac{TF} model \cite{Colin2000} is used to resolve the reactive zone of the flames computing the reaction rates issuing from the \ac{ARC} mechanism directly on the grid. Its dynamic formulation is used here \cite{Legier2000}: the correction of the \ac{TF} model is applied only where a reactive zone is detected by a flame sensor modeled as a function of the fuel consumption rate \cite{Jaravel2017}. The thickening factor is adapted to the local size of the cells in the mesh. 
Unresolved flame wrinkling effects are accounted for using an efficiency function based on the wrinkling factor of Charlette et al. \cite{Charlette2002}. 
Flame thickening is not triggered in the discharge region so that it does not influence the ignition induced by the \acp{NRPD}. To ensure a correct resolution of the physical phenomena related to the \acp{NRPD}, a very fine mesh is used between the discharge zone and the outlet of the \acf{SB}. This approach has already proven to be able to account for plasma physics in the same combustor \cite{Male2022}. Schulz et al. have also validated the modeling approach in a similar sequential combustor \cite{Schulz2019a}.

The effects of the plasma described by the phenomenological model detailed in Section~\ref{sec:plasma_modeling} are integrated into the \ac{LES} balance equations. Since plasma chemistry is much faster than flow mixing, its subgrid-scale interaction with turbulence is neglected \cite{Bechane2020}. 
The filtered species mass balance equations read
\begin{equation}
    \frac{\partial \bar{\rho} \widetilde{Y_k}}{\partial t} + \frac{\partial}{\partial x_j}\left(\bar{\rho} \widetilde{Y_k} \widetilde{u_j}\right)
    = -\frac{\partial}{\partial x_j} \left[ \overline{J_{j, k}} + {\overline{J_{j, k}}^t} \right]
    + \frac{\mathcal{E}}{\mathcal{F}} \dot{\omega}_k^c + \dot{\omega}_k^p  \;
    \text { for } k=1, N_s \, \text{,}
\end{equation}
where $\bar{\varphi}$ and $\widetilde{\varphi}$ are the Reynolds and Favre filtering of a variable $\varphi$. $\rho$ is the gas density, $Y_k$ the mass fraction of the species $k$, and $u_j$ the velocity along the direction $j$. $\overline{J_{j, k}}$ is the filtered diffusive flux and $\overline{J_{j, k}}^t$ the subgrid-scale flux of the species $k$. $\mathcal{E}$ and $\mathcal{F}$ are the efficiency and the thickening factor of the \ac{TF} model, respectively. $\dot{\omega}_k^c$ and $\dot{\omega}_k^p$ are the source term of the species $k$ due to combustion and plasma kinetics, respectively. 
The balance equation for the total non-chemical energy $E$, sum of the sensible and kinetic energy, reads
\begin{equation}
    \frac{\partial \bar{\rho} \widetilde{E}}{\partial t}+\frac{\partial}{\partial x_j}\left(\bar{\rho} \widetilde{E} \widetilde{u_j}\right)
    = - \frac{\partial}{\partial x_j} \left[ \overline{u_i \left(P \delta_{i j}-\tau_{i j}\right)} + \overline{q_j} + {\overline{q_j}^t} \right] 
    + \frac{\mathcal{E}}{\mathcal{F}} \dot{\omega}_T + {\dot{E}}_{\mathrm{heat}}^p + {\dot{R}}_{\mathrm{VT}}^p
    \, \text{,}
    \label{eq:LES_nrj}
\end{equation}
where $P$ is the pressure and $\delta_{ij}$ the Kronecker symbol. $\tau_{i j}$ is the viscous tensor. $\overline{q_j}$ is the filtered energy flux and ${\overline{q_j}^t}$ the subgrid-scale energy flux in the direction $j$. $\dot{\omega}_T$ is the \ac{HRR} due to combustion. 
The filtered momentum balance equations are unchanged. 
In addition, a balance equation is solved for the out-of-equilibrium vibrational energy $e_{\mathrm{vib}}$:
\begin{equation}
    \frac{\partial \bar{\rho} \widetilde{e_{\mathrm{vib}}}}{\partial t} + \frac{\partial}{\partial x_j}\left(\bar{\rho} \widetilde{e_{\mathrm{vib}}} \widetilde{u_j}\right)
    = \left(\left[\frac{\bar{\mu}}{S_{c, e_{\mathrm{vib}} }}+\frac{\mu_t}{S_{c, e_{\mathrm{vib}} }^t}\right] \frac{\partial \widetilde{e_{\mathrm{vib}}}}{\partial x_i}\right)
    + {\dot{E}}_{\mathrm{vib}}^p - {\dot{R}}_{\mathrm{VT}}^p
    \, \text{,}
    \label{eq:LES_evib}
\end{equation}
where a classical Fick's law is used for the molecular and turbulent diffusion. $\mu$ is the dynamic viscosity and $S_c$ the Schmidt number. It is assumed that $e_{\mathrm{vib}}$ diffuses as N\textsubscript{2} (i.e., ${S_{c, e_{\mathrm{vib}} }} = {S_{c, {\mathrm{N_2}} }}$) which are the dominant vibrationally excited molecules. 
The relaxation term ${\dot{R}}_{\mathrm{VT}}^p$ that appears in Eqs~(\ref{eq:LES_nrj}) and (\ref{eq:LES_evib}) controls the rate at which the out-of-equilibrium vibrational energy is converted to translational energy (i.e., gas heating). 
This relaxation is dominated by 
\reaction{N_2(v=1) + M -> N_2 (v=0) + M + heat \text{,} \label{ceq:RVT}}
that slowly heats the gas. The rate of Eq.~(\ref{ceq:RVT}) strongly depends on the colliding partner $\mathrm{M}$ for which N\textsubscript{2}, O\textsubscript{2}, CO\textsubscript{2}, H\textsubscript{2}O, H\textsubscript{2}, OH, H and O are considered. 
The relaxation term ${\dot{R}}_{\mathrm{VT}}^p$ is computed using the Landau-Teller harmonic oscillator assumption \cite{Landau_1936,capitelli2011plasma}:
\begin{equation}
    {\dot{R}}_{\mathrm{VT}}^p = \rho \frac{ e_{\mathrm{vib}} }{ \tau_{\mathrm{VT}} } \; \text{with} \, \frac{1}{\tau_{\mathrm{VT}}} = \sum_\mathrm{M} \frac{1}{\tau_{\mathrm{VT}}^{\mathrm{M}}} \, \text{,}
\end{equation}
where $\tau_{\mathrm{VT}}$ is a relaxation time computed from ${\tau_{\mathrm{VT}}^{\mathrm{M}}}$ using coefficients from Refs~\cite{Starikovskiy2013,Millikan1963} for the processes given by Eq.~(\ref{ceq:RVT}).

The discharge power distribution $\dot{E}^p(x,y,z,t)$ follows a spatial density function $\mathcal{F}_v (x,y,z)$ and is assumed to be temporally constant during the discharge duration $T_p$:
\begin{equation}
    \dot{E}^p(x, y, z, t)= \begin{cases}\frac{E^p}{T_p} \mathcal{F}_v (x, y, z) & \text { if } t^{\prime} \in\left[0, T_p\right] \\ 0 & \text { otherwise}\end{cases} \; \text{with} \int_V \mathcal{F}_v dV = 1 \,    \text{,}
\end{equation}
where $t' = t \; \mathrm{mod} \left( 1 / \mathrm{PRF} \right)$ is defined as the time from the beginning of the current pulse. The discharge duration $T_p$ includes the high voltage pulse during which the discharge energy is deposited, and the relaxation of the electronic states. It is set to $50$~ns as in Refs~\cite{Castela2016,Barleon2023,Male2022,Bechane2020,Bechane2022}. The spatial distribution of the plasma power $\mathcal{F}_v (x,y,z)$ is estimated as a cylinder with a radius of $500$~$\mathrm{\mu m}$ and a length of $5$~mm in the $y$-direction between the pin-to-pin electrodes. 
The shape and size of the discharge are obtained from plasma channel imaging (Section~\ref{sec:seqcombconfig}).

\begin{figure}[hbt!]
    \centering
    \includegraphics[width=1.0\textwidth]{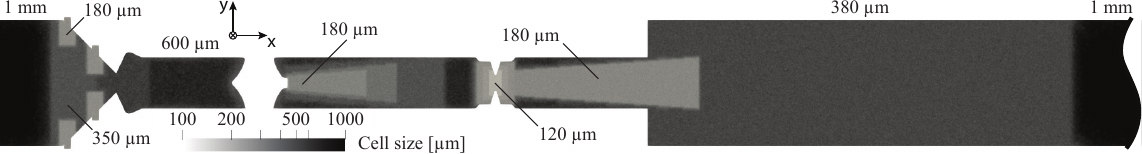}
    \caption{Planar cut of the mesh colored by the cell size.}
    \label{fig:mesh_view}
\end{figure}

The numerical domain is discretized with an unstructured mesh comprised of 146 million tetrahedral elements (Fig.~\ref{fig:mesh_view}). 
The size of the cells is adapted according to the regions of interest. Refinement zones are applied in the dilution air mixer and after the fuel injection nozzle. Special attention is given to the discharge area where the mesh size is 120 $\mathrm{\mu m}$. In addition, the region between the electrodes and the combustion chamber is discretized with a mesh size of 180 $\mathrm{\mu m}$ to properly capture the effects of NRPDs \cite{Male2022}. The combustion chamber is meshed with a cell size of 380 $\mathrm{\mu m}$, resulting in a flame thickening factor $\mathcal{F} \approx 5$. 
Walls are handled with a law-of-the-wall formulation \cite{Schmitt2007}. They are modeled as isothermal with a temperature of $900$ and $700$ K for the quartz and water-cooled aluminium parts, respectively. The quartz parts are the side walls of the sequential burner and the combustion chamber, designed for optical access. 
Navier–Stokes characteristic boundary conditions \cite{Poinsot1992} are used for the inlets and the outlet of the domain. 

%

\FloatBarrier
\section{Results and discussion}
\label{sec:discussion}

\subsection{Characterization of the discharges}\label{sec:discharge_characterization}

Detailed kinetic simulations are performed in \ac{0D} isochoric reactors using the plasma-combustion chemical mechanism developed in Ref.~\cite{Cheng2022} to characterize the discharge effects. The mechanism has been extensively validated against experimental measurements of \ac{NEP} discharges in air and methane-air mixtures. The measurements include: atomic oxygen production \cite{Rusterholtz2013,Uddi2009}, fast gas heating \cite{Rusterholtz2013,Montello2013}, vibrationally excited nitrogen population \cite{Montello2013}, and ignition delay time \cite{Aleksandrov2009a}. The mechanism has been later extended to mixtures containing combustion products in Ref.~\cite{Barleon:phd2022}. Cantera \cite{cantera} is coupled with the \acf{EEDF} solver BOLOS \cite{Hagelaar2005} to perform \ac{0D} simulations including the detailed plasma-combustion kinetics \cite{Cheng2022}. The volume of the discharge is experimentally estimated as a cylinder with a radius of $500$~$\mathrm{\mu m}$ and a length of $5$~mm (Section~\ref{sec:seqcombconfig}). The energy density $e^p = E^p / V_d$, where $V_d$ is the discharge volume, is $e^p = 0.308$~$\mathrm{J/cm^3}$ for $E^p = 1.21$~mJ, and $e^p = 0.527$~$\mathrm{J/cm^3}$ for $E^p = 2.07$~mJ. 

\begin{figure}[hbt!]
    \centering
    \includegraphics[width=0.45\textwidth]{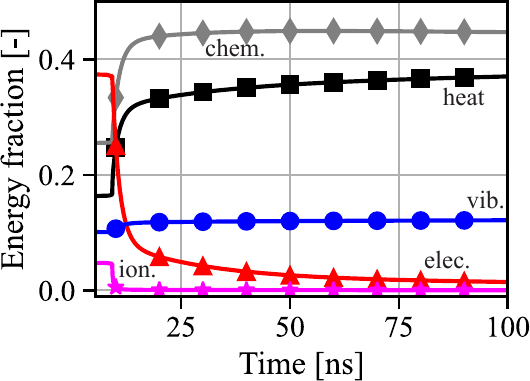}
    \caption{Distribution of the discharge energy with $E/N = 246$~Td, $e^p = 0.308$~$\mathrm{J/cm^3}$, 
    and the initial conditions given in Table~\ref{tab:compo_dloc}.}
    \label{fig:energy_fraction_246Td_1p21mJ}
\end{figure}

In order to determine an initial state for the discharge, a first \ac{LES} is performed on the \ac{OP} ``NRPD off'' (Table~\ref{tab:parameters}). The thermochemical state variables inside the discharge volume are then spatially and temporally averaged (Table~\ref{tab:compo_dloc}). From this thermochemical state, a high voltage pulse is applied until the deposited energy reaches the deposited energy measured in the experiments. The reduced electric field is estimated as $E/N = U / d / N \approx 246$~Td, where $U = 14$~kV is the applied voltage, $d = 8$~mm is the interelectrode gap distance and $N = 7.12 \times 10^{24}$~$\mathrm{m^{-3}}$ is the gas number density. Then, the simulation is continued up to $100$~ns in order to relax the electronic excitation energy into other energy channels. Figure~\ref{fig:energy_fraction_246Td_1p21mJ} shows the distribution of the discharge energy for the $E^p = 1.21$~mJ ($e^p = 0.308$~$\mathrm{J/cm^3}$) case. The ionization energy quickly vanishes after the high voltage pulse. The electronic excitation energy relaxes during the $100$~ns to reach a negligible residue as in Ref~\cite{Barleon2023}. At time $t = 100$~ns, the energy fractions $\alpha_{\mathrm{chem}}$, $\alpha_{\mathrm{heat}}$ and $\alpha_{\mathrm{vib}}$ are $47.5$~\%, $39.9$~\% and $12.6$~\%, respectively. The $\alpha$ coefficients for the plasma modeling are detailed in Table~\ref{tab:alpha}. The specific chemical $\alpha_{i=1,...,J}$ for the $J$ processes detailed in Table~\ref{tab:PACMIND} are determined using the averaged molar production rates from the detailed plasma kinetic simulation \cite{Barleon2023}. Similar energy distribution is observed for $E^p = 2.07$~mJ since energy branching is governed by $E/N$. 

\begin{table}[hbt!]
\centering
\small
\begin{tabular}{|l|c|l|c|}
\hline
$T$ {[}K{]}         & 1030.6 &                     &       \\ \hline
$X_{\mathrm{O_2}}$  & 0.116  & $X_{\mathrm{H_2}}$  & 0.045 \\ \hline
$X_{\mathrm{N_2}}$  & 0.689  & $X_{\mathrm{CO_2}}$ & 0.033 \\ \hline
$X_{\mathrm{CH_4}}$ & 0.049  & $X_{\mathrm{H_2O}}$ & 0.068 \\ \hline
\end{tabular}
\caption{Thermochemical state averaged in space and time in the discharge volume. Extract from the \ac{LES} computation ``NRPD off''. Only the species with a molar fraction $X_k \geq 10^{-3}$ are reported.}
\label{tab:compo_dloc}
\end{table}

\begin{table}[hbt!]
\centering
\small
\begin{tabular}{|l|c|l|c|}
\hline
$\alpha_{\mathrm{chem}}$ & 0.475 & $\alpha_{3}$ & 0.005 \\ \hline
$\alpha_{\mathrm{heat}}$ & 0.399 & $\alpha_{4}$ & 0.076 \\ \hline
$\alpha_{\mathrm{vib}}$  & 0.126 & $\alpha_{5}$ & 0.026 \\ \hline
$\alpha_{1}$             & 0.183 & $\alpha_{6}$ & 0.134 \\ \hline
$\alpha_{2}$             & 0.049 & $\alpha_{7}$ & 0.000 \\ \hline
\end{tabular}
\caption{$\alpha$ coefficients for the phenomenological plasma model computed from a detailed kinetic plasma simulation with $E/N = 246$~Td, $e^p = 0.308$~$\mathrm{J/cm^3}$, 
and the initial conditions given in Table~\ref{tab:compo_dloc}.}
\label{tab:alpha}
\end{table}

\begin{figure}[hbt!]
    \centering
    \includegraphics[width=0.4\textwidth]{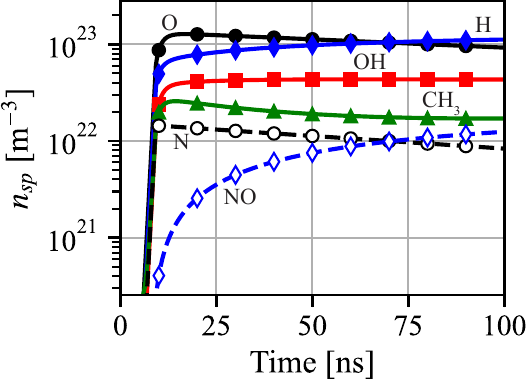}
    \caption{Number density of the main species produced by the \ac{NEP} discharge with $E/N = 246$~Td, $e^p = 0.308$~$\mathrm{J/cm^3}$, 
    and the initial conditions given in Table~\ref{tab:compo_dloc}.}
    \label{fig:species_density_246Td_1p21mJ}
\end{figure}

\begin{figure}[hbt!]
    \centering
    \includegraphics[width=0.5\textwidth]{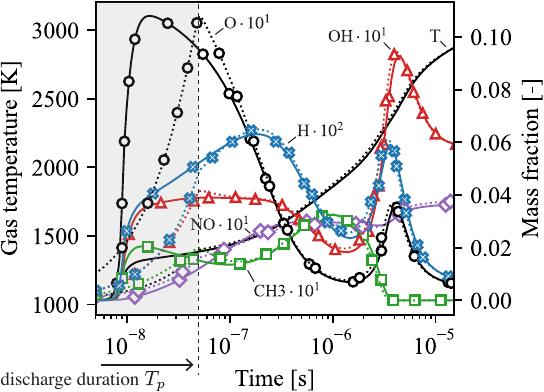}
    \caption{Gas temperature and species mass fraction after a \ac{NEP} discharge using detailed plasma chemistry (\full{}) and the plasma modeling approach (\dotted{}). $E/N = 246$~Td, $e^p = 0.308$~$\mathrm{J/cm^3}$, 
    initial conditions given in Table~\ref{tab:compo_dloc}.}
    \label{fig:246Td_1p21mJ_temperature_mfrac}
\end{figure}

O, H and OH are the main species quickly produced by the discharge as shown in Fig.~\ref{fig:species_density_246Td_1p21mJ}. NO is not abruptly produced during the first $10$~ns. It is produced later in time due to chemical reactions that will be detailed in Section~\ref{sec:nox_chemistry}.
The results of the detailed kinetics are compared with those from the phenomenological model in a similar isochoric \ac{0D} reactor (Fig.~\ref{fig:246Td_1p21mJ_temperature_mfrac}) using Cantera \cite{cantera}, where the source terms from the plasma model are simply added to the \ac{RHS} of the governing equations. The modeling of the plasma effects does not consider the different phases of the plasma observed during the high voltage pulse and the afterglow. Thus, differences in species concentration are observed up to $t = T_p = 50$~ns. Afterwards, the evolution of the species and the ignition of the reactive mixture are very well retrieved. This validates the approach used in this work to model the effects of the \ac{NEP}.

\subsection{Flame response to NRPDs}\label{sec:flame_response}

\begin{figure}[hbt!]
    \centering
    \includegraphics[width=1.0\textwidth]{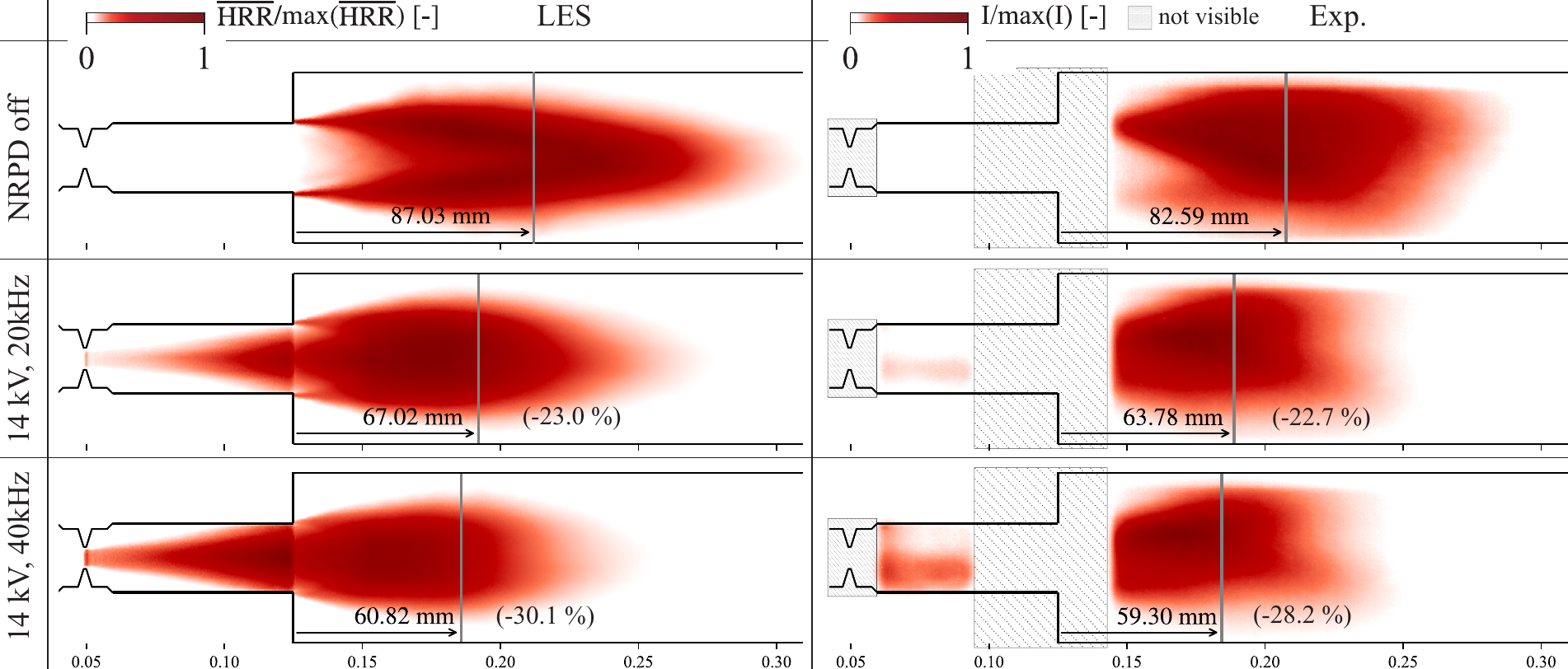}
    \caption{\ac{LOS} integration of the average \ac{HRR} from the \acp{LES} (left) and of the average \OHstar{} chemiluminescence from the experiments (right) at three different plasma parameters. All values are normalized with respect to their maximum value. The flame center of gravity $Z_{\mathrm{COG}}$ (Eq.~(\ref{eq:ZCOG})) is indicated by a vertical gray line. Areas without visual access are indicated by hatched boxes.}
    \label{fig:ZCOG_ave}
\end{figure}

The effect of the \acp{NRPD} on the average flame position is quantified by the position of the flame \ac{COG} as depicted in Fig.~\ref{fig:ZCOG_ave}. The flame \ac{COG} is estimated using the \ac{LOS} integration of the \ac{HRR} for the \acp{LES} and the \OHstar{} chemiluminescence\textemdash used as a tracer of \ac{HRR} \cite{Lee2003}\textemdash for the experiments. 
The distance from the \ac{COG} to the \acf{SB} outlet in the $z$-direction is defined as
\begin{equation}
    Z_{\mathrm{COG}} = \frac{\sum I_i \, \left(z_i - z_{\mathrm{SB}} \right) }{\sum I_i} \, \text{,}
    \label{eq:ZCOG}
\end{equation}
where $I_i$ is the \ac{HRR} or \OHstar{} chemiluminescence intensity of the $i$\textsuperscript{th} data point, and $z_i$ its $z$-coordinate. The \ac{LES} data found in the non-visible parts of the experiments (hatched boxes in Fig.~\ref{fig:ZCOG_ave}) are not taken into account so as not to bias the comparison. 
The moving back of the flame center of gravity following the application of the \acp{NRPD} is well retrieved by the \ac{LES}. The absolute value of $Z_{\mathrm{COG}}$ differs by a few millimetres. This may be due to the experimental configuration which produces an asymmetrical flame, particularly noticeable for ``NRPD off'' in Fig.~\ref{fig:ZCOG_ave}. The asymmetry is due to the design of the injector, which does not produce a perfect distribution of the fuel within the \ac{SB}. 
Moreover, it is important to bear in mind that in non-perfectly premixed flames, the flame \ac{HRR} and chemiluminescence are not directly proportional and the later is only an approximate marker of the former, the reliability of the approximation degrading with the degree of non-premixedness. 
Experimental images also show weak (``14kV, 20kHz'') and intense (``14kV, 40kHz'') chemical activity in the SB before the combustion chamber. This is also retrieved by the LES. 
The capability of the \ac{LES} setup to capture the extension of the lean blow-off limit by \acp{NRPD} has already been demonstrated by the authors \cite{Male2022}. Here, it is shown that the numerical approach also reproduces the effect of plasma on the flame position. This reinforces the conclusions in Ref.~\cite{Male2022} regarding the capability of the LES strategy to retrieve the effects of the plasma on complex combustor configurations. 

\FloatBarrier
\subsection{Nitrogen oxides chemistry}
\label{sec:nox_chemistry}

\subsubsection{3D LES results}\label{sec:3DLESNOX}

\begin{figure}[hbt!]
    \centering
    \includegraphics[width=0.5\textwidth]{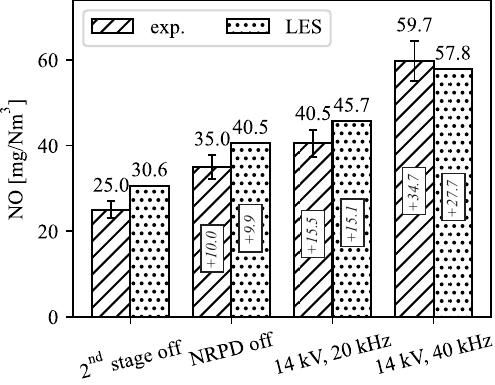}
    \caption{Average NO mass concentration at the probe location. Gas without \water{} and cooled to $276.15$~K. The stickers show the increment compared to ``2\textsuperscript{nd} stage off'' where the 2\textsuperscript{nd} stage fuel injection and the plasma is off. Table~\ref{tab:parameters} characterizes the four different \acp{OP}.}
    \label{fig:NOX_values}
\end{figure}

The NO concentration experimentally measured at the probe location indicated in Fig.~\ref{fig:sketch_LES_seqcomb} (``Gas analyzer probe'') is compared with the one from the \ac{LES} to validate the modeling. The NO concentration is extracted from the \ac{LES} on a plane normal to the $z$ axis at the position of the experimental probe. First, the thermochemical state of the mixture is averaged over this plane and over time to reproduce an experimental sample. Then, \water{} is removed from the gas composition and the mixture is cooled to $276.15$~K to match the conditioning of the gas sample by the gas analyzer. 
The absolute values obtained using \ac{LES} are in very good agreement with the experimental ones (Fig.~\ref{fig:NOX_values}). The NO increase due to the 2\textsuperscript{nd} stage combustion of $+10$~$\mathrm{mg/Nm^3}$ is perfectly retrieved (``NRPD off''). The NO increase due to plasma at ``$14$~kV, $20$~kHz'' ($+15.5$~$\mathrm{mg/Nm^3}$) is also well captured. A slight deviation is observed for ``$14$~kV, $40$~kHz'' where the NO increment is $+27.7$ instead of $+34.7$~$\mathrm{mg/Nm^3}$ for the experiments. 
Overall, the increase in NO is very well retrieved by the numerical modeling. This confirms the capability of the LES setup to reproduce the effects of the \acp{NRPD} on the NO chemistry.

\begin{figure}[p]
    \centering
    \includegraphics[width=1.0\textwidth]{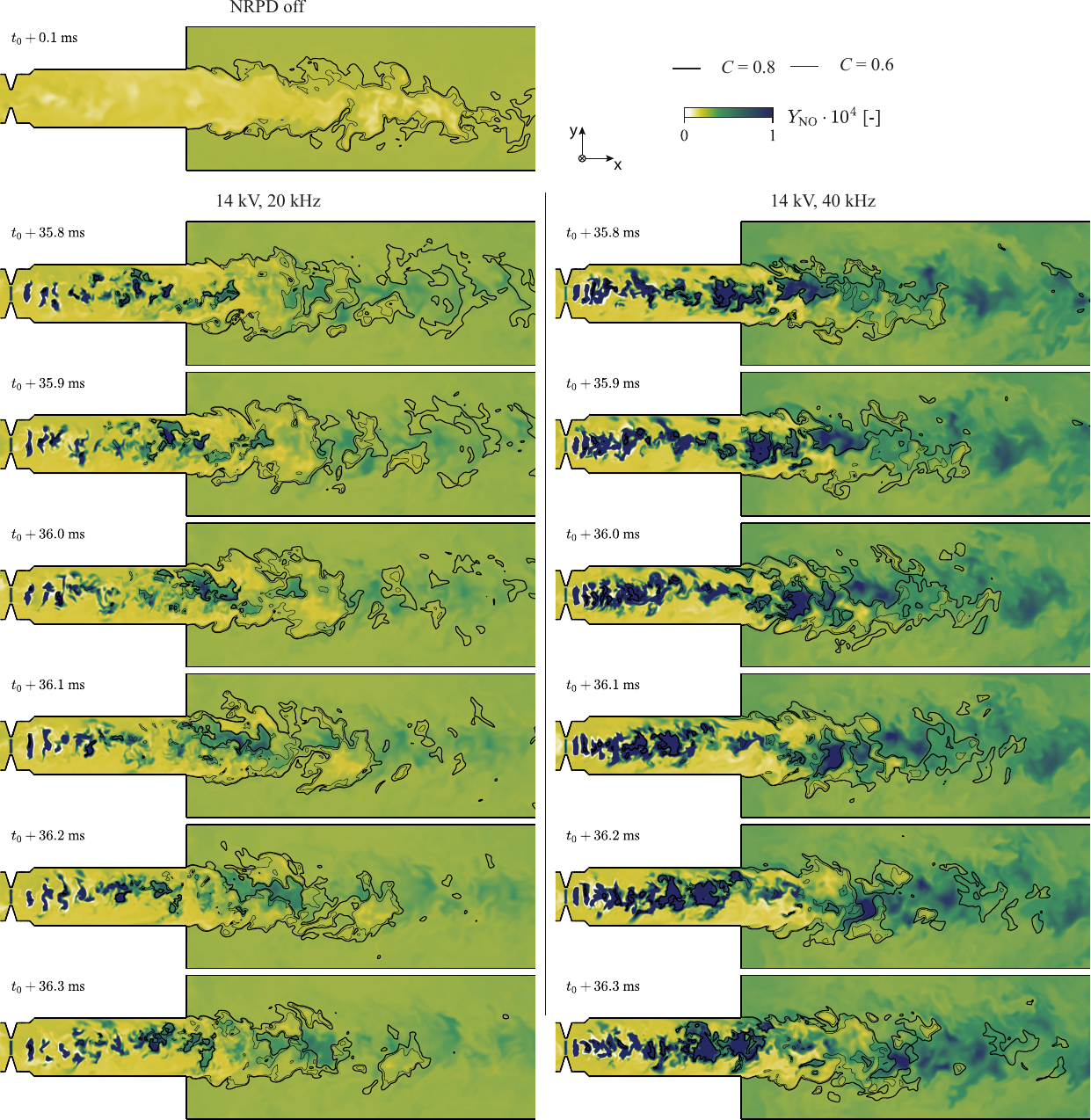}
    \caption{Planar cuts colored by the NO mass fraction with two iso-contours of progress variable $C$ (Eq.~(\ref{eq:progress-variable})) for three different \acp{OP}.}
    \label{fig:YNO}
\end{figure}

\begin{figure}[hbt!]
    \centering
    \includegraphics[width=0.60\textwidth]{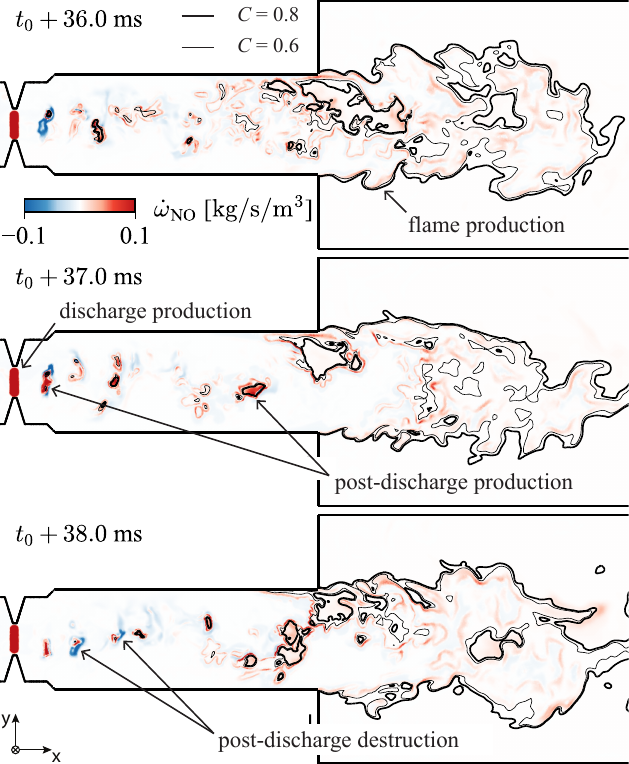}
    \caption{Planar cuts colored by the NO chemical source term with two iso-contours of progress variable $C$ (Eq.~(\ref{eq:progress-variable})). ``$14$~kV, $20$~kHz''.}
    \label{fig:wNO}
\end{figure}

Figure~\ref{fig:YNO} shows the NO mass fraction in the \ac{LES} domain for the different \acp{OP}. The flame front is delineated with two iso-contours of progress variable $C$ defined as
\begin{equation}
    C = \frac{Y_C-Y_C^u}{Y_C^b-Y_C^u} \, \text{,}
    \label{eq:progress-variable}
\end{equation}
with
\begin{equation}
    Y_C = Y_\mathrm{H_2O} + Y_\mathrm{CO_2} \, \text{,}
\end{equation}
where the $u$ and $b$ superscripts represent the value in the unburnt and burnt gas, respectively. NO is produced right after the discharges in the kernels generated by the plasma. These kernels with high NO concentration are convected to the flame brush while distorted by and diffusing in the turbulent flow. The chemical activity within the turbulent flame in the combustion chamber does not heavily affect the NO molecules that simply flow toward the exhaust. This is also seen in Fig.~\ref{fig:wNO} where NO consumption is limited within the flame fronts. However, there are zones where NO is heavily consumed after the discharge location indicated as ``post-discharge destruction'' in Fig.~\ref{fig:wNO}. They follow the intense NO production by the plasma in the discharge volume (``discharge production'' in Fig.~\ref{fig:wNO}). In addition, zones with intense NO production are also seen after the discharge, in the \ac{SB} (``post-discharge production'' in Fig.~\ref{fig:wNO}). 

%
%

\FloatBarrier
\subsubsection{Zero-dimensional study of the detailed chemistry}\label{sec:0D_detailed_plasma_chem_NO}

\begin{figure}[hbt!]
    \centering
    \includegraphics[width=0.50\textwidth]{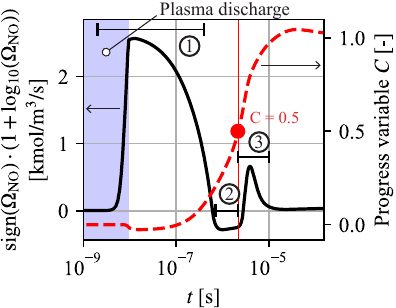}
    \caption{Magnitude of the NO molar source term $\Omega_{\mathrm{NO}}$ after a \ac{NEP} discharge using detailed plasma chemistry in a \ac{0D} plasma reactor with $E/N = 246$~Td, $e^p = 0.308$~$\mathrm{J/cm^3}$, 
    initial conditions given in Table~\ref{tab:compo_dloc}. The blue area represent the discharge period during which the electric field is applied. The progress variable $C$ (Eq.~(\ref{eq:progress-variable})) shows the combustion advancement. (1) NO plasma production, (2) NO consumption, (3) post-discharge NO production.}
    \label{fig:omega_NO}
\end{figure}

An analysis of the plasma-combustion chemistry is required to understand the kinetic pathways leading to the NO production and destruction zones. Due to the complexity of the kinetics, a \ac{0D} approach is used where the plasma reactor described in Section~\ref{sec:discharge_characterization} is run for a longer time. For the sake of brevity, only the \ac{OP} ``$14$~kV, $20$~kHz'' is analyzed. The dominant reactions are the same for ``$14$~kV, $40$~kHz''.
Three different phases for NO production/destruction are highlighted by the molar NO source term $\Omega_{\mathrm{NO}}$ in Fig.~\ref{fig:omega_NO}:
\begin{itemize}
    \item[(1)] High NO production due to the plasma discharge. The effects of the plasma (e.g., gas heating, radical production) generate NO intensively for a short period of time. The production of NO during this phase is $30.6 \times 10^{-3}$~$\mathrm{mol/m^3}$.
    \item[(2)] Post-discharge NO destruction. Later in time, NO is actually consumed while combustion is occurring. The consumption of NO during this phase is $1.1 \times 10^{-3}$~$\mathrm{mol/m^3}$.
    \item[(3)] Post-discharge NO production. During the combustion after $C\approx0.5$, NO production rises again before fading at the end of the combustion process. The production of NO during this phase is $7.5 \times 10^{-3}$~$\mathrm{mol/m^3}$. This is four times less than the NO production during (1).
\end{itemize}

\begin{figure}[hbt!]
    \centering
    \includegraphics[width=1.00\textwidth]{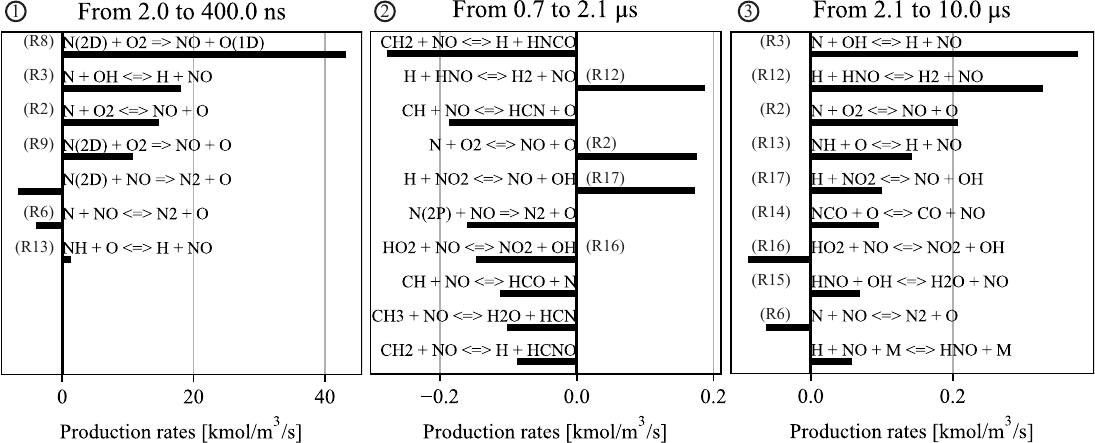}
    \caption{Averaged NO production rates for the three different periods annotated in Fig.~\ref{fig:omega_NO}. Only the highest fluxes are shown. $E/N = 246$~Td, $e^p = 0.308$~$\mathrm{J/cm^3}$, 
    initial conditions given in Table~\ref{tab:compo_dloc}.}
    \label{fig:prodRates_NO}
\end{figure}

The NO production rates due to the specific chemical reactions in these three phases are shown in Fig.~\ref{fig:prodRates_NO}. Plasma NO production in phase (1) is mainly due to the presence of atomic N and N(2D), which react with \dioxygen{} and OH to form NO molecules via the reactions:
\reaction{N(2D) + O2 -> NO + O(1D) \text{,} \label{ceq:N(2D)+O2->NO+O(1D)}}
\[ \ce{ N + OH <=> H + NO \text{,} }\tag{\ref{ceq:N+OH}} \]
\[ \ce{ N + O2 <=> NO + O \text{ and} }\tag{\ref{ceq:N+O2}} \]
\reaction{N(2D) + O2 -> NO + O \text{.} \label{ceq:N(2D)+O2<=>NO+O}} 
For this case, N and N(2D) are mainly produced by the electron impact reaction
\reaction{e^- + N2 -> e^- + N + N(2D) \text{.} \label{ceq:E+N2->E+N+N(2D)}} 
One might have expected the reaction
\[ \ce{ N2 + O <=> NO + N \text{} }\tag{\ref{ceq:N2+O}} \]
to be a source of NO production because of the atomic oxygen produced by the plasma (Fig.~\ref{fig:species_density_246Td_1p21mJ}). However, the backward reaction rate dominates the forward reaction rate in Eq.~(\ref{ceq:N2+O}) during this period. Thereby, NO is effectively consumed by this reaction.


The NO consumption observed in phase (2) is mainly due to ``reburn'' mechanisms where NO reacts with hydrocarbons and is subsequently reduced via 
\reaction{CH_i + NO -> products \text{,} \label{ceq:reburn}}
with $\mathrm{i}=1,2,3$. This logically coincides with the start of the combustion, when \methane{} is decomposed via the oxidation process. 

After the fuel oxidation phase during the combustion process (phase (3)), N atoms still play a role via the Eqs~(\ref{ceq:N+O2}) and (\ref{ceq:N+OH}). In addition, reactions part of the so-called ``prompt'' mechanism \cite{Lamoureux2010} play a role in the NO production. They are mainly:
\reaction{H + HNO <=> H2 + NO \text{,} \label{ceq:prompt_H+HNO}}
\reaction{NH + O <=> H + NO \text{,} \label{ceq:prompt_NH+O}}
\reaction{NCO + O <=> CO + NO \text{ and} \label{ceq:prompt_NCO+O}}
\reaction{HNO + OH <=> H2O + NO \text{.} \label{ceq:prompt_HNO+OH}}

\FloatBarrier
\subsubsection{Study of the LES chemistry}
The NO chemistry is now studied in the complex 3D turbulent environment of the sequential burner from the \ac{LES} for the \ac{OP} ``$14$~kV, $20$~kHz''. In order to identify the reactions responsible for NO production and consumption, an index of production $I_{\mathrm{NO}}^{P,j}$ and of consumption $I_{\mathrm{NO}}^{C,j}$ are defined for every reaction $j$. They read
\begin{equation}
    I_{\mathrm{NO}}^{P,j} = \frac{ \nu_{\mathrm{NO}}^j Q^j  }{ \Omega_{\mathrm{NO}}^P  }  \, \text{and}
    \label{eq:NO_production_index}
\end{equation}
\begin{equation}
    I_{\mathrm{NO}}^{C,j} = \frac{ \nu_{\mathrm{NO}}^j Q^j  }{ \Omega_{\mathrm{NO}}^C  }  \, \text{,}
    \label{eq:NO_consumption_index}
\end{equation}
where $\nu_{\mathrm{NO}}^j$ and $Q^j$ are the stoichiometric coefficient of NO and the rate of progress of the reaction $j$, respectively. $\Omega_{\mathrm{NO}}^P$ and $\Omega_{\mathrm{NO}}^C$ represent the production and consumption rate of NO and are defined as
\begin{equation}
    \Omega_{\mathrm{NO}}^P = \sum_j \mathrm{max} \left(  \nu_{\mathrm{NO}}^j Q^j, 0 \right) \, \text{and}
\end{equation}
\begin{equation}
    \Omega_{\mathrm{NO}}^C = \sum_j \mathrm{min} \left(  \nu_{\mathrm{NO}}^j Q^j, 0 \right) \, \text{.}
\end{equation}

\begin{figure}[hbt!]
    \centering
    \includegraphics[width=1.00\textwidth]{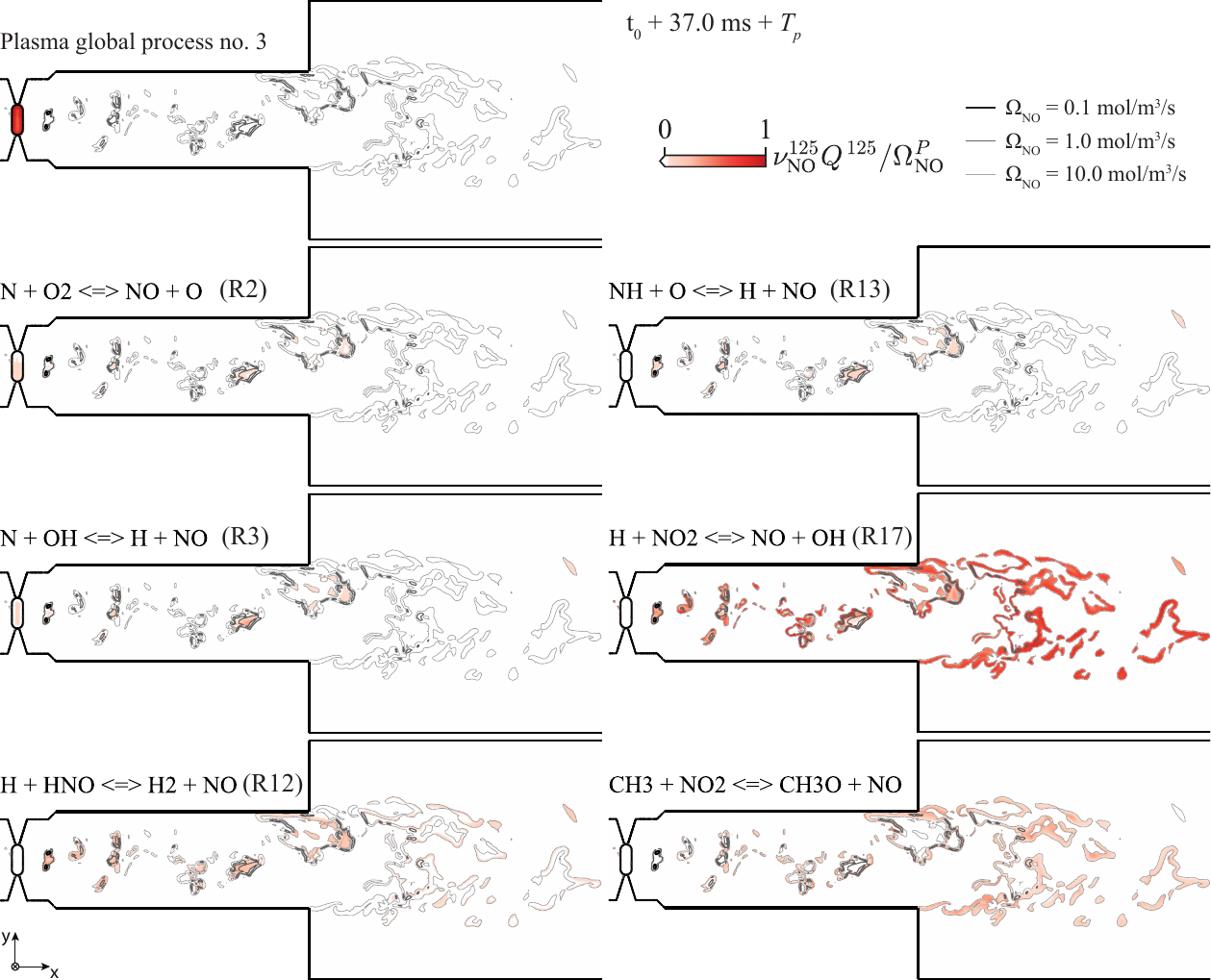}
    \caption{Planar cuts colored by the NO production index (Eq.~(\ref{eq:NO_production_index})) with three iso-contours of NO molar source term $\Omega_{\mathrm{NO}}$. ``$14$~kV, $20$~kHz''. Snapshot at the end of the occurrence of a \ac{NEP} discharge.}
    \label{fig:production_14kv_20kHz_37ms}
\end{figure}

Figure~\ref{fig:production_14kv_20kHz_37ms} shows the most important NO production indexes on a plane through the \acf{SB} at the end of the occurrence of a \ac{NEP} discharge. The plasma global process no. 3 (Table~\ref{tab:PACMIND}) is the main source of NO in the discharge zone. The reactions (\ref{ceq:N+O2}) and (\ref{ceq:N+OH}) play a secondary role in this region. The species generated by the plasma (Table~\ref{tab:PACMIND}) provide a favorable environment for the development of these two reactions. 
Reactions (\ref{ceq:prompt_H+HNO}) and (\ref{ceq:prompt_NH+O}) from the prompt NO mechanism play a role later in the igniting kernels as in Fig.~\ref{fig:prodRates_NO}. A conversion of \noo{}, another component of nitrogen oxides, to NO is also observed both in the plasma-induced igniting kernels and in the main flame brush.

\begin{figure}[hbt!]
    \centering
    \includegraphics[width=0.5\textwidth]{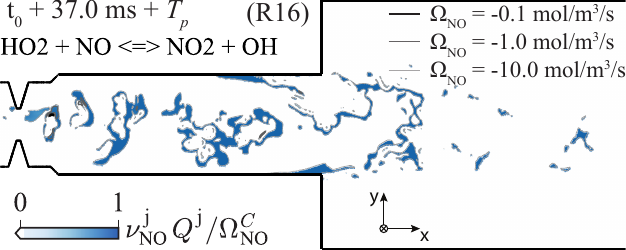}
    \caption{Planar cut colored by the NO consumption index (Eq.~(\ref{eq:NO_consumption_index})) with three iso-contours of NO molar source term $\Omega_{\mathrm{NO}}$. ``$14$~kV, $20$~kHz''. Snapshot at the end of the occurrence of a \ac{NEP} discharge.}
    \label{fig:destruction_14kv_20kHz_37ms}
\end{figure}

The destruction/consumption of NO in the \ac{SB} after the discharge location is mainly due to a single reaction (Fig.~\ref{fig:destruction_14kv_20kHz_37ms}):
\reaction{HO2 + NO <=> NO2 + OH \text{.} \label{ceq:HO2+NO<=>NO2+OH}}
This reaction takes place next to the NO production zones, on the fresh gas side. This is due to the diffusion of NO molecules into the fresh gas containing HO\textsubscript{2} radicals. Hydroperoxyl is strongly present in the hot mixture of vitiated air and fuel that flows through the \ac{SB} and undergoes the first stages of the auto-ignition process. This cannot be retrieved by the \ac{0D} study performed in Section~\ref{sec:0D_detailed_plasma_chem_NO} because no exchange with the surrounding environment is considered. Reburn processes (Eq.~(\ref{ceq:reburn})) do not effectively participate in NO depletion in the turbulent flow. This is due to the low magnitude of the associated source term shown in Fig.~\ref{fig:omega_NO}, phase (2).

\FloatBarrier
\subsubsection{Impact of the nitrogen plasma chemistry}

\begin{figure}[hbt!]
    \centering
    \includegraphics[width=1.00\textwidth]{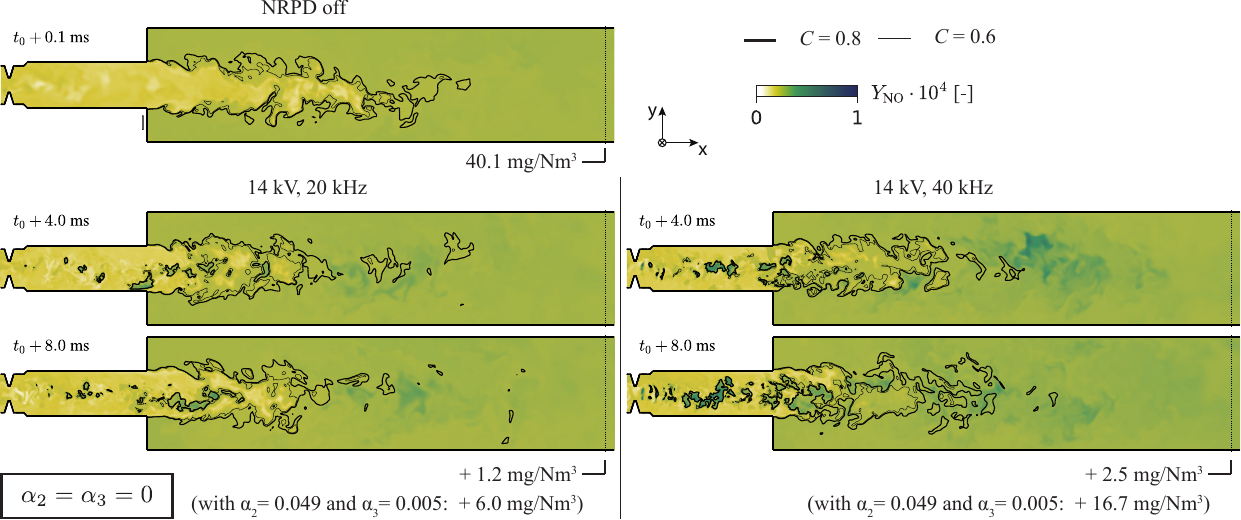}
    \caption{Planar cuts colored by the NO mass fraction with two iso-contours of progress variable $C$ (Eq.~(\ref{eq:progress-variable})). The nitrogen chemistry is removed from the plasma modeling (i.e., $\alpha_2=\alpha_3=0$ in Table~\ref{tab:alpha}). The NO concentration is extracted from a plane normal to the $z$ axis, delineated by a dotted line, $250$~mm after the inlet of the 2\textsuperscript{nd} stage combustion chamber.}
    \label{fig:YNO_woAlphaNOx}
\end{figure}

To evaluate the impact of the nitrogen plasma chemistry on the NO production the coefficient $\alpha_2$ and $\alpha_3$ are set to zero. These coefficients are responsible for the production of N and NO in the phenomenological plasma model employed (Table~\ref{tab:PACMIND}). Figure~\ref{fig:YNO_woAlphaNOx} shows the mass fraction of NO significantly lower than the one obtained taking into account the nitrogen plasma chemistry (Fig.~\ref{fig:YNO}). To limit the computational time, the concentration is extracted $250$~mm after the inlet of the 2\textsuperscript{nd} stage combustion chamber, rather than $450$~mm where the experimental probe is located. This is possible here because no comparison with experimental data is made and the \ac{LES} setup has already been validated in Sections \ref{sec:flame_response} and \ref{sec:3DLESNOX}. The NO concentration (gas without \water{} and cooled to $276.15$~K) increases only slightly when $\alpha_2=\alpha_3=0$ ($+0.4$~$\mathrm{mg/Nm^3}$ and $+2.5$~$\mathrm{mg/Nm^3}$ for ``$14$~kV, $20$~kHz'' and ``$14$~kV, $40$~kHz'', respectively), whereas it increases significantly when plasma nitrogen chemistry is considered ($+6.0$~$\mathrm{mg/Nm^3}$ and $+16.7$~$\mathrm{mg/Nm^3}$ for ``$14$~kV, $20$~kHz'' and ``$14$~kV, $40$~kHz'', respectively) compared to the case without NRPD. This demonstrates the importance of taking into account nitrogen plasma chemistry for the correct prediction of NO emissions during plasma assisted combustion.

\begin{figure}[p]
    \centering
    \includegraphics[width=1.00\textwidth]{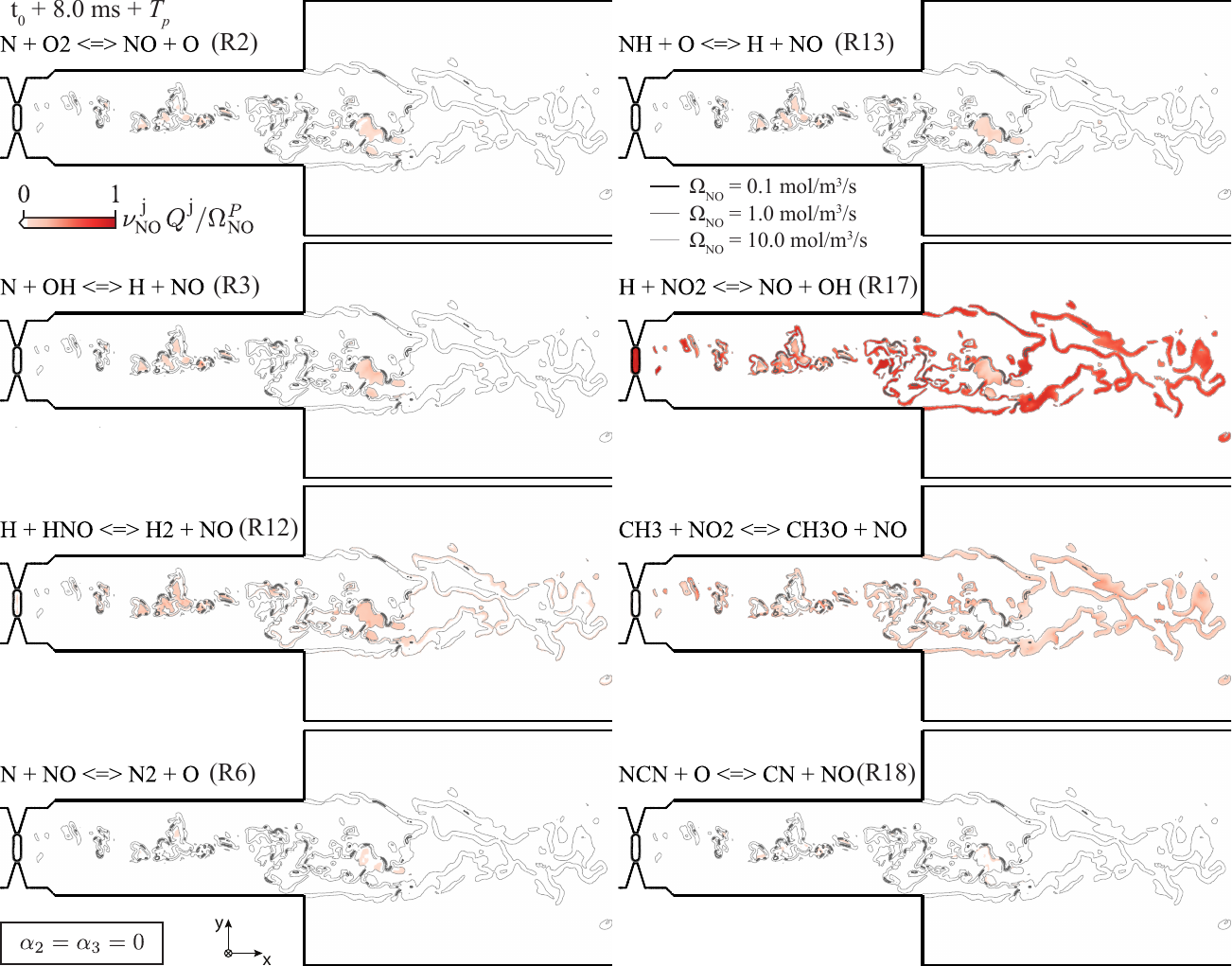}
    \caption{Planar cuts colored by the NO production index (Eq.~(\ref{eq:NO_production_index})) with three iso-contours of NO molar source term $\Omega_{\mathrm{NO}}$. ``$14$~kV, $20$~kHz''. The nitrogen chemistry is removed from the plasma modeling (i.e., $\alpha_2=\alpha_3=0$ in Table~\ref{tab:alpha}). Snapshot at the end of the occurrence of a \ac{NEP} discharge.}
    \label{fig:production_14kv_20kHz_woAlphaNOx_8ms}
\end{figure}

The NO production index (Eq.~(\ref{eq:NO_production_index})) of the most important reactions for NO production is shown in Fig.~\ref{fig:production_14kv_20kHz_woAlphaNOx_8ms}. Since the production of atomic N by the plasma is disabled, the reactions (\ref{ceq:N+O2}) and (\ref{ceq:N+OH}) no longer take place in the discharge zone. They are delayed and are now taking place in the igniting kernels together with the initiation reaction of the thermal NO pathway defined by Zel'dovich
\[ \ce{ N2 + O <=> NO + N \text{.} }\tag{\ref{ceq:N2+O}} \]
The generation of NO due to the plasma-generated species in the discharge zone is now mainly due to the presence of H atoms induced by the processes 4 and 6 in Table~\ref{tab:PACMIND}. \noo{} from the 1\textsuperscript{st} stage combustion reacts with H to produce NO via
\reaction{H + NO2 <=> NO + OH \text{.} \label{ceq:H+NO2<=>NO+OH}}
Prompt mechanism still plays a role with the reactions (\ref{ceq:prompt_H+HNO}), (\ref{ceq:prompt_NH+O}) and 
\reaction{NCN + O <=> CN + NO \text{.} \label{ceq:NCN+O<=>CN+NO}}
Similar results are found for ``$14$~kV, $40$~kHz''.

\FloatBarrier
\section{Conclusion}
\label{sec:conclusion}

The formation of NO during plasma assisted combustion in a lab-scale sequential combustor was investigated using numerical simulation tools. 
The complex 3D turbulent reactive flow has been simulated with a \ac{LES} approach, including a precise description of the chemical kinetics and an accurate modeling of the plasma effects. Simulation results were found to be in very good agreement with experimental measurements, including \OHstar{} chemiluminescence imaging and NO sampling at the burner outlet. The 3D analysis is complemented by detailed \ac{0D} plasma computations. 

For the operating points studied, the NO formation is mainly due to the dissociation of \dinitrogen{} by direct electron impact. This generates nitrogen atoms that react with OH and \dioxygen{} to form NO in the sequential burner. Within the flame brush, the prompt mechanism slightly participates to the NO formation, without drastically changing the concentration. 
Removing the modeling of the nitrogen plasma chemistry from the description of the plasma effects leads to a strong underestimation of the NO formation. This shows the limited participation of the thermal routes in the formation of NO. In addition, this demonstrates the need for an advanced modeling approach of the \ac{NEP} in order to study the formation of undesirable species such as NO. 

More generally, this study is the first to identify the NO formation pathways when \acp{NRPD} are applied in a turbulent-flow combustor. The excellent agreement with experiments demonstrates the capability of the \ac{LES} setup to retrieve NRPD effects on combustion and NO formation. 
This work demonstrates how leading-edge simulation tools can help research on innovative combustion technology. 

To optimize the application of NRPDs to combustion systems, particular attention needs to be paid to the reactions causing \dinitrogen{} dissociation, found to be mainly responsible for the formation of NO. Work is underway to identify the NRPD key parameters that trigger the dissociation of nitrogen molecules, in order to give design indications for practical applications. 
It is possible that the contribution of kinetic pathways to \nox{} formation will change for other \acp{OP}, especially at higher equivalence ratio or operating pressure. Experiments at higher pressures are currently in progress in our laboratory to investigate this open question.


%

\section*{Acknowledgement}

This project has received funding from the European Research Council (ERC) under the European Union’s Horizon 2020 research and innovation programme (grant agreement No [820091]).
This work was supported by a grant from the Swiss National Supercomputing Centre (CSCS) under project ID s1220.
The authors gratefully acknowledge CERFACS for providing the LES solver AVBP and the ARCANE library. They especially thank Q. Cazères and J. Wirtz for their help in using ARCANE.

%

\FloatBarrier
 \bibliographystyle{elsarticle-num} 
 \bibliography{library,custom}





\end{document}